\documentclass[prl,twocolumn,showpacs,nofootinbib]{revtex4}

\usepackage{graphicx}
\usepackage{dcolumn}
\usepackage{bm}

\begin{document}

\preprint{APS/A. W. Holleitner}

\title{Suppression of Spin Relaxation in Submicron InGaAs Wires}

\author{A. W. Holleitner\footnote{present address: Center for NanoScience (CeNS), Munich,
Germany}}
\author{V. Sih}
\author{R. C. Myers}
\author{A. C. Gossard}
\author{D. D. Awschalom}
\email{awsch@physics.ucsb.edu}
\affiliation{%
Center for Spintronics and Quantum Computation, University of
California, Santa Barbara, CA 93106, USA}

\date{\today}

\begin{abstract}
We investigate electron spin dynamics in narrow two-dimensional
$n$-InGaAs channels as a function of the channel width. The spin
relaxation times increase with decreasing channel width, in
accordance with recent theoretical predictions based on the
dimensionally-constrained D'yakonov-Perel' mechanism.
Surprisingly, the suppression of the relaxation rate, which is
anticipated for the one-dimensional limit, is observed for widths
that are an order of magnitude larger than the electron mean free
path. We find the spin precession length and the channel width to
be the relevant length scales for interpreting these results.

\end{abstract}


\maketitle

In the emerging field of spintronics, it is important to explore
carrier spin relaxation mechanisms in nanostructures as a function
of dimensionality. The effect of reducing feature sizes in
spintronic devices is relevant for future technological
applications~\cite{wolf01,awsch02}. In two and three dimensions,
elementary rotations do not commute, with significant impact on
the spin dynamics if the spin precession is induced by spin-orbit
coupling~\cite{kato04}. Spin-orbit coupling creates a randomizing
momentum-dependent effective magnetic field; the corresponding
relaxation process is known as the D'yakonov-Perel' (DP)
mechanism~\cite{dyakonov}. In an ideal one-dimensional system,
however, all spin rotations are limited to a single axis, and the
spin rotation operators commute. In the regime approaching the
one-dimensional limit, a progressive slowing and finally a
complete suppression of the DP spin relaxation have been
predicted, if the lateral width of a two-dimensional channel is
reduced to be on the order of the electron mean free
path~\cite{bournel,malshukov,kiselev,pareek}. The predictions are
made for semiconductor heterostructures, such as InGaAs quantum
wells, in which the spin-orbit interactions are dominated by
structural inversion asymmetry
(SIA)~\cite{nitta,koga,grundler,ganichev}. Such solid-state
systems have been proposed as candidates for spintronic devices,
including spin transistors~\cite{datta}, due to their potential
scalability and compatibility with existing semiconductor
technology.

Here, we combine optical time-resolved Faraday rotation (TRFR)
spectroscopy with magnetotransport measurements in
two-dimensional, n-doped InGaAs quantum well channels. As a
function of the channel width, we extract the spin relaxation time
and the elastic scattering times of the electrons. Surprisingly,
experiments on wide channels, with widths of an order of magnitude
larger than the electron mean free path $l_{e}$, reveal an
effective slowing of the spin relaxation. In this regime, the data
show that the spin relaxation is dominated by the DP mechanism.
For narrower channels, we find that an interplay between the spin
precession length $l_{SP}$ and the channel width $w$ determines
the electron spin dynamics in the wires. A saturation of the
slowing spin relaxation is found for the narrowest wires,
indicating other sources of spin relaxation exist such as the
cubic spin-orbit coupling term due to bulk inversion asymmetry
(BIA)~\cite{winkler} and the spin relaxation mechanism proposed by
Elliot and Yafet~\cite{elliott}.

The spin splitting in a two-dimensional quantum well due to SIA
can be expressed in the form of an effective angular frequency
vector
\begin{eqnarray} {\bf \Omega}({\bf k}) = (1\slash l_{SP})[{\bf v}({\bf k}) \times \hat{z}]
\end{eqnarray}
with {\bf k} the momentum vector, and {\bf v}({\bf k}) the
velocity of an electron~\cite{awsch02}. $\hat{z}$ is the unit
vector perpendicular to the quantum well, and $l_{SP}$ is the spin
precession length, over which the electrons remain spin polarized.
Given a system with a fixed mean free path, a larger effective
angular frequency induces faster spin rotations and, in turn, a
shorter spin relaxation time. In the case of motional
narrowing~\cite{lau}, the corresponding spin relaxation rate can
be described as
\begin{eqnarray} \tau_{SP}^{-1} = |{\bf \Omega}({\bf k})|^{2} \tau_{M} / 2
\end{eqnarray}
with $\tau_{M}$ the momentum scattering time. In order to probe
the spin dynamics for different momentum vectors, transport and
spin coherence experiments are performed on a set of n-doped
InGaAs wires [Fig.~1(a)]. Wires are patterned along the
crystallographic directions [100], [110], [010] and
[$\overline{1}$10], while the spins are optically oriented along
the growth direction [001]. Structures are fabricated by e-beam
lithography and reactive ion etching out of three modulation-doped
$n$-In$_{0.2}$Ga$_{0.8}$As/GaAs quantum wells. The unpatterned
quantum wells A, B, and C have the following sheet densities
$n_{s}$ and mobilities at a temperature of $T$ = 5 K: (A) 5.4
$\times$ 10$^{11}$ cm$^{-2}$ and 3.8 $\times$ 10$^{4}$ cm$^{2}$/V
s, (B) 6.6 $\times$ 10$^{11}$ cm$^{-2}$ and 3.1 $\times$ 10$^{4}$
cm$^{2}$/V s, and (C) 7.0 $\times$ 10$^{11}$ cm$^{-2}$ and 2.4
$\times$ 10$^{4}$ cm$^{2}$/V s. The quantum wells are situated 100
nm below the surface of the heterostructures, and the quantum well
width is $\Delta$z = 7.5 nm (for more details on the growth, see
Ref.~\cite{sih}). The widths of the wires $w$ range between 420 nm
and 20 $\mu$m, and the height of the wires is chosen to be 150 nm
[Fig.~1(b)]. For the optical experiments, the wires are arranged
in arrays with the dimension of 200 $\times$ 200 $\mu$m$^{2}$,
while the diameter of the laser spot is about 50 $\mu$m. In order
to provide constant etching parameters for all widths and
directions of the wires, the distance between adjacent wires is
set to be 1 $\mu$m for all of the arrays. Magnetotransport
experiments are performed on single wires fabricated with the same
etching parameters~\cite{wires}.

\begin{figure}
\includegraphics[width=0.44\textwidth]{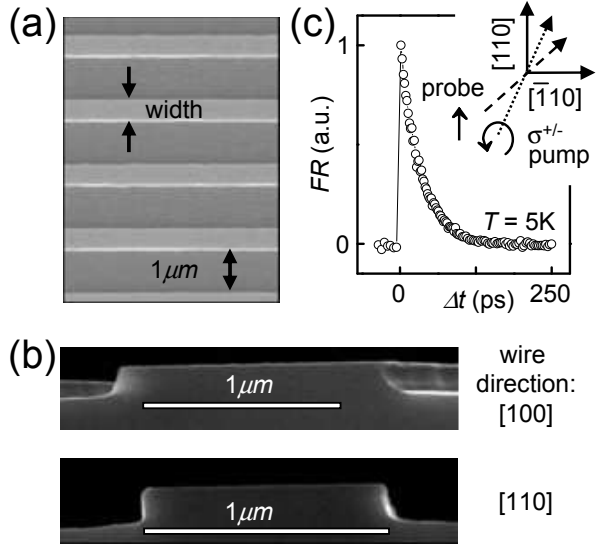}
\caption{\label{fig:epsart} (a) Scanning electron micrograph (SEM)
of dry-etched InGaAs wires, which are patterned along the four
crystallographic directions [100], [110], [010], and
[$\overline{1}$10]. The wire widths are varied between 420 nm and
20 $\mu$m, while their separation is fixed at 1 $\mu$m. (b) SEM of
Sample A along the [$\overline{1}$10] cleaving direction. Both
wires along [100] and [110] have a width of (1.02 $\pm$ 0.04)
$\mu$m. (c) Time-resolved Faraday rotation (TRFR) of 750 nm wires
patterned along [010] on Sample A at zero magnetic field. Inset: a
circularly-polarized pump pulse excites the spin polarization. A
time-delayed linearly-polarized pulse probes the spin dynamics.}
\end{figure}

The electron spin dynamics are probed with the TRFR technique,
using 100 fs pulse trains from a mode-locked Ti:Sapphire laser
tuned to the absorption edge of the quantum wells, $E_{LASER}$ =
1.37 eV [Fig. 1(c)]~\cite{sih}. The evolution of the Faraday
rotation angle can be described by a single exponential decay
$\Theta_{F} = A_{1}e^{\Delta t/\tau_{SP}}$, where $A_{1}$ is the
amplitude of the electron-spin polarization and $\Delta$t is the
time delay between the circularly-polarized pump and the
linearly-polarized probe pulse. As shown with solid lines in
Fig.~2(a), the exponential behavior of the data is described by a
longitudinal spin relaxation time $\tau_{SP}$ for both the
unpatterned quantum well (open squares) and for the wires aligned
along different crystallographic directions (data for Sample B at
5 K)~\cite{T1}. For all samples, we find that at widths narrower
than $\sim$10 $\mu$m, the spin relaxation times in the wires are
longer than in the unpatterned quantum wells [Fig.~2(b)]. In
addition, we find that wires aligned along [100] and [010] show
equivalent spin relaxation times, which are generally longer than
the spin relaxation times of wires patterned along [110] and
[$\overline{1}$10] (for clarity, only the data for the directions
[100] and [110] are shown). All data are obtained by measuring the
transmission signal in the Faraday geometry.

\begin{figure}
\includegraphics[width=0.44\textwidth]{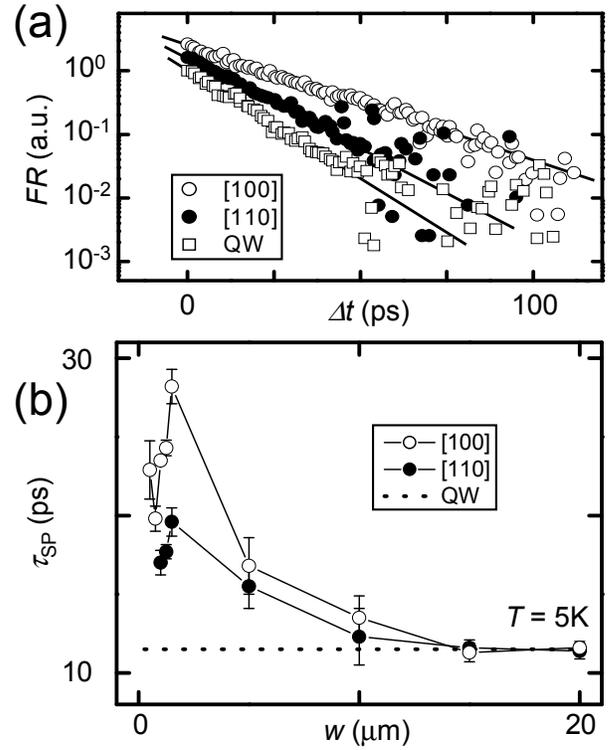}
\caption{\label{fig:epsart} (a) Faraday rotation at 5 K for Sample
B (open squares) and 750 nm wires patterned along [100] (open
circles) and [110] (filled circles). Black lines are guides to the
eye, and the data are off-set for clarity. (b) Width dependence of
spin relaxation times for wires fabricated from Sample C. The
dotted line depicts the spin relaxation time of the unpatterned
quantum well. Measurements were performed at B~=~0.}
\end{figure}

\begin{figure}
\includegraphics[width=0.44\textwidth]{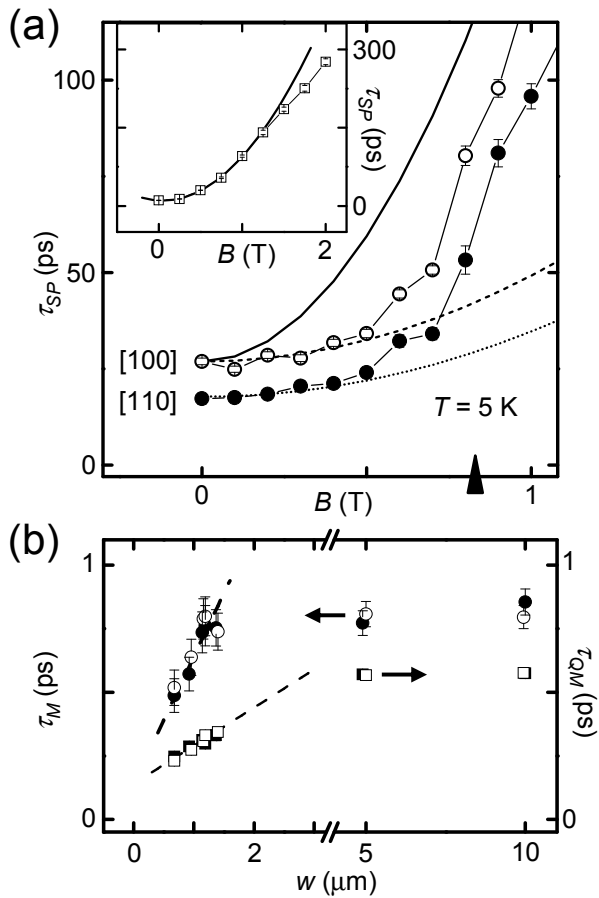}
\caption{\label{fig:epsart} (a) Magnetic field dependence of spin
relaxation times in the unpatterned Sample C (inset) and wires
(open and filled circles for [100] and [110], respectively). The
magnetic field is applied perpendicular to the surface of the
sample. The dotted, dashed, and black lines are fits to Eq.~(3),
and the triangle indicates $B_{QM}$ (see text for details). (b)
Momentum scattering time (circles) and quantum lifetime (squares)
versus channel width for the directions [100] (open symbols) and
[110] (filled symbols), respectively. The dashed lines are guides
to the eye. }
\end{figure}

If an external magnetic field is applied perpendicular to the
quantum wells, the precession axis of the electron spin can be
fixed independently of the scattered momentum vector. In the case
that the DP mechanism is the dominant relaxation process, the
following magnetic field dependence of the spin relaxation time
has been predicted (for $\omega_{C}\tau$ $<$ 1)~\cite{optical}:
\begin{eqnarray} \tau_{SP}(B) = \tau_{SP} (0) [1 +
(\omega_{C}\tau)^{2}]
\end{eqnarray}
where $\omega_{C}$ = $eB/m^{*}$ is the cyclotron frequency of an
electron with charge $e$, $m^{*}$ = 0.064 m$_{e}$ is the effective
electron mass~\cite{sih}, and $\tau$ represents the intrinsic
elastic scattering time. The magnetic field dependence of the spin
relaxation time for the unpatterned quantum wells is well fit by
this prediction [Fig.~3(a), inset]. We find that $\tau$ $\sim$ 1
ps, in agreement with the measured momentum scattering time
$\tau_{M}$ in these quantum wells~\cite{sih}. Figure~3(a) displays
$\tau_{SP}$ as a function of magnetic field for wires with $w$ =
1.25 $\mu$m patterned on wafer C. The solid line shows the
prediction according to Eq.~(3) with a momentum scattering time of
$\tau_{M}$ = (7.6 $\pm$ 0.2) $\times$ 10$^{13}$ s. In addition, we
determine an estimate of the quantum lifetime $\tau_{QM}$ through
magnetotransport measurements on single wires by plotting the
Shubnikov-deHaas oscillations in a Dingle plot~\cite{dingle}.
Surprisingly, the optical data is better fit using the quantum
lifetime $\tau_{QM}$ = (3.1 $\pm$ 0.1) $\times$ 10$^{-13}$ s
(dashed and dotted lines for the directions [100] and [110],
respectively). The condition $\omega_{C}\tau_{QM} = 1$ can be
represented by a magnetic field $B_{QM} = m^{*}/e \tau_{QM}$,
which is depicted as a triangle. It can be nicely seen that
Eq.~(3) describes the data well for $\omega_{C}\tau_{QM} < 1$.
This field-dependence of $\tau_{SP}$ suggests that (i) the DP
mechanism is indeed the dominant spin relaxation mechanism in the
studied structures and (ii) the quantum lifetime $\tau_{QM}$ is
the relevant time scale for the wires at low magnetic field values
where the Zeeman energy is negligible.

Figure~3(b) shows the dependence of $\tau_{M}$ and $\tau_{QM}$ on
the channel width. Both scattering times show a rapid decrease for
the narrowest channels. Since the spin relaxation times greatly
exceed the charge scattering times, the quantum wells can be
considered to be in the "motional narrowing" regime
[Eq.~(2)]~\cite{lau}. Figure~3(b) further demonstrates that
$\tau_{M}$ is constant for wires with $w \geq 1.2$ $\mu$m,
independent of the crystallographic direction (the value of the
momentum scattering time corresponds to a mean free path $l_{e}$ =
(275 $\pm$ 5) nm). In Fig.~2(b), however, we find an enhanced spin
coherence for wires with $w \leq 5$ $\mu$m. This implies that for
$1.2$ $\mu$m $\leq w \leq 5$ $\mu$m, the effective angular
frequency $|{\bf \Omega}({\bf k})|$ is reduced, according to
Eq.~(2). At the same time, we observe from the images shown in
Fig.~1(b) that the channels are homogeneously etched.
Consequently, strain relaxation in the quantum wells via
dislocation nucleation is unlikely for wires with $1.2$ $\mu$m
$\leq w \leq 5$ $\mu$m and a quantum well width of $\Delta$z = 7.5
nm~\cite{knotz}. This interpretation is supported by the fact that
the two-dimensional electron density $n_{S}$ in the channels shows
no dependence on the channel width and direction down to $w \sim$
400 nm.

\begin{figure}
\includegraphics[width=0.44\textwidth]{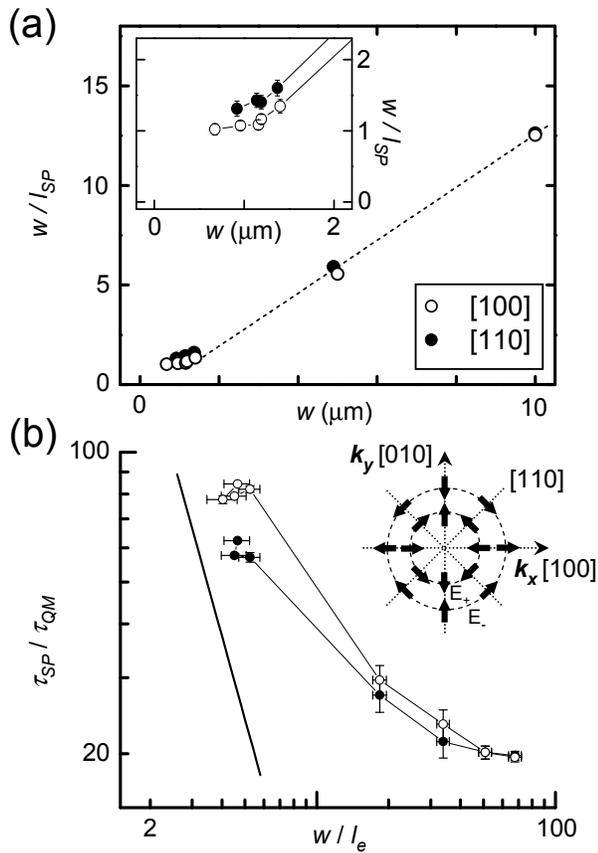}
\caption{\label{fig:epsart} (a) Ratio of the channel width $w$ and
the spin diffusion length $l_{SP}$ as a function of $w$. Inset:
For narrow wires, the channel boundaries limit the spin diffusion
length. (b) Logarithmic presentation of the spin relaxation time
in units of the scattering time and the mean free length. Black
line depicts the quasi one-dimensional limit (open and closed
circles for [100] and [110], respectively). Inset: Schematic
vector map of the spin eigenfunctions in a quantum well with bulk
inversion asymmetry.}
\end{figure}

Figure~4(a) shows the ratio between the wire width $w$ and the
spin diffusion length
\begin{eqnarray}
l_{SD} = \sqrt{\tau_{SP} \cdot v_{F}^{2} \tau_{M} /2}
\end{eqnarray}
as a function of $w$. In the motional narrowing regime, the spin
diffusion length is the same as the spin precession length;
inserting Eq.~(2) into Eq.~(4) yields $l_{SD} = v_{F}/|{\bf
\Omega(k)}| = l_{SP}$. For wide channels, the spin
precession/diffusion length is given by the two-dimensional limit,
i.e. a linear dependence of $w/l_{SP}$ versus $w$ (dashed line).
For narrow widths, however, the data suggest that the spin
diffusion length is ultimately limited by the wire width
($w/l_{SP}$ $\sim$ 1) [Fig.~4(a), inset]. Concerning the spin
diffusion length, the narrowest wires act as quasi one-dimensional
channels and, in turn, the two-dimensional spin dynamics are
constrained by the side walls of the
wires~\cite{bournel,malshukov,kiselev,pareek}. Figure~4(b) depicts
the ratio of $\tau_{SP}$ and $\tau_{QM}$ as a function of the mean
free path $l_{e}$ in a logarithmic scale. A qualitatively similar
graph is obtained using $\tau_{M}$ instead of $\tau_{QM}$. The
graph closely resembles the predictions of Ref.~\cite{kiselev},
which indicates that the slowing of the spin relaxation in the
wires is due to a dimensionally-constrained DP mechanism, as
predicted for SIA~\cite{bournel,malshukov,kiselev,pareek}. The DP
spin relaxation due to BIA eventually limits this
slowing~\cite{malshukov}. An anisotropy in the spin-splitting and,
thus, in $|{\bf \Omega}({\bf k})|$ has been predicted for InGaAs
quantum wells assuming cubic BIA terms and Fermi wave vectors
which are comparable to $k_{F}$ = $\sqrt{2\pi n_{S}} \cong (0.018
- 0.021)$ \.{A}$^{-1}$ of the discussed samples~\cite{winkler}.
Since the spin splitting due to BIA is anisotropic, the magnitude
of $|{\bf \Omega(k)}|$ depends sensitively on the momentum vector.
This explains why spin lifetimes are similar for channels oriented
along [100] and [010], but are different from wires patterned
along the [110] and [$\overline{1}$10] directions. The inset of
Fig.~4(b) depicts the orientation of the spin eigenfunctions for
two spin-split subbands E$_{+}$ and E$_{-}$ of a zincblende
quantum well in the presence of BIA (E$_{+}$ and E$_{-}$ are
defined as in Ref.~\cite{winkler}). For SIA, however, Eq. (1)
suggests a constant value of $|{\bf \Omega(k)}|$ that only depends
on the magnitude of {\bf k} and which is always oriented
perpendicular to {\bf k}. For the narrowest channels, the data do
not reach the predicted behavior of $\tau_{SP} \sim w^{-2}$ [black
line in Fig.~4(b)], where the channel width would limit the mean
free path~\cite{malshukov,kiselev}. Instead, we find a saturation
of the spin relaxation time. Since the Elliott-Yafet mechanism
becomes more effective for shorter scattering times, this
relaxation mechanism ultimately limits the slow-down of the spin
relaxation in the narrowest channels~\cite{elliott}. At the same
time, a negligible pump power dependence of the TRFR data supports
the interpretation that the spin relaxation mechanism proposed by
Bir, Aronov, and Pikus is only of minor importance to the spin
dynamics in the InGaAs wires~\cite{bir}.

Generally, we utilize InGaAs quantum wells with a relatively low
In concentration and an electron mobility $\mu \cong(2-4)\times
10^{4}$ cm$^{2}$/V s. The spin precession length $l_{SP}
\cong(0.9-1.1)\mu$m [Fig.~4(a)] yields a Rashba spin coupling
constant of $\alpha \equiv \hbar^{2}/(2 l_{SP} m^{\star}) \cong
(0.5-0.7) \times 10^{-12}$ eV m~\cite{awsch02,kiselev}, in good
agreement with previous results on InGaAs quantum
wells~\cite{koga,das}. This set of parameters ensures that the
quantum wells are in the "motional narrowing" regime, in order to
detect the dimensionally-constrained DP
mechanism~\cite{bournel,malshukov,kiselev,pareek}. Coupling
constants of $\alpha \cong 4 \times 10^{-11}$ eV m have been
achieved by increasing the In concentration in the quantum
wells~\cite{nitta,grundler,datta,das}. Larger coupling constants
entail relatively short spin precession lengths and thus shorter
spin relaxation times, which can be compensated by lowering the
electron mobility.

In summary, an effective slowing of the D'yakonov-Perel' (DP) spin
relaxation mechanism is observed in unexpectedly wide conducting
channels of $n$-InGaAs quantum wires. The results are consistent
with a dimensionally-constrained DP mechanism as recently
predicted for narrow two-dimensional quantum wells exhibiting
structural inversion asymmetry. For the narrowest wires with only
a few hundreds of nanometers width, an interplay between the spin
diffusion length and the wire width determines the spin dynamics.\\

\begin{acknowledgments}
We thank Y. Li for technical support and F. Meier, V. Khrapay and
J. P. Kotthaus for stimulating discussions. We gratefully
acknowledge financial support by the AFOSR, DMEA, NSF and ONR.
\end{acknowledgments}


\begin{thebibliography}{99}
\bibitem{wires} After patterning, $n_{s}$ is reduced by about
(17$\pm$1)\%, and the mobility is lowered by about (5$\pm$5)\%.
\bibitem{T1} The longitudinal spin relaxation time is often denoted as 'T$_{1}$'.
\end{thebibliography}
\end{document}